\documentclass[oneside,reqno,10pt]{amsart}

\usepackage{graphicx}
\usepackage{epstopdf}
\usepackage{amsmath}
\usepackage{amssymb}
\usepackage[top=1in, bottom=1.20in, left=1.25in, right=1.25in]{geometry}
\usepackage{textcomp}
\usepackage{mathrsfs}
\usepackage{tikz}
\usepackage{tikz-cd}
\usepackage[utf8]{inputenc}
\usepackage{mathabx}
\usepackage{bm}
\usepackage{amsthm}
\usepackage{amsfonts}
\usepackage[colorlinks=true,linkcolor=blue]{hyperref}
\usepackage{pb-diagram}
\usepackage{lipsum}
\usepackage{secdot} 
\usepackage{color}
\usepackage[normalem]{ulem}
\usepackage{enumitem}
\usepackage{dsfont}
\usepackage{appendix} 
\usepackage{color}
\usepackage{calc}
\usepackage{slantsc}
\usepackage[sc]{mathpazo}
\usepackage{indentfirst}
\usepackage{mathtools}
\usepackage{hyperref}
\usepackage{amscd}
\usepackage{caption}
\usepackage{subcaption}
\usepackage{fancybox}
\usepackage{wrapfig}
\usepackage[all]{xy}
\usepackage{verbatim}
\usepackage{stmaryrd}

\theoremstyle{plain}
\newtheorem{theorem}{Theorem}[section]

\newtheorem{defn}[theorem]{Definition}
\newtheorem{example}[theorem]{Example}
\newtheorem{remark}[theorem]{Remark}

\numberwithin{equation}{section}

\newcommand{\twobar}{/\kern-0.5em/}
\newcommand{\threebar}{/\kern-0.5em/\kern-0.5em/}

\newcommand{\Z}{\mathbb{Z}}

\newcommand{\R}{\mathbb{R}}

\newcommand{\cV}{\mathcal{V}}

\newcommand{\Hom}{\mathrm{Hom}}

\newcommand{\End}{\mathrm{End}}

\newcommand{\GL}{\mathrm{GL}}

\newcommand{\cA}{\mathcal{A}}

\begin{document}
	
\title{Quantum Finite Automata and Quiver Algebras}

\author{George Jeffreys}
\address{Department of Mathematics and Statistics, Boston University, 111 Cummington Mall, Boston MA 02215, USA}
\email{georgej@bu.edu}

\author{Siu-Cheong Lau}
\address{Department of Mathematics and Statistics, Boston University, 111 Cummington Mall, Boston MA 02215, USA}
\email{lau@math.bu.edu}

\maketitle

\begin{abstract}
We find an application in quantum finite automata for the ideas and results of [JL21] and [JL22]. We reformulate quantum finite automata with multiple-time measurements using the algebraic notion of near-ring.  This gives a unified understanding towards quantum computing and deep learning.  When the near-ring comes from a quiver, we have a nice moduli space of computing machines with metric that can be optimized by gradient descent.
\end{abstract}

\section{Motivation: QFA and Near-Ring}

In quantum theory, evolution of states is unitary.  An observable is modeled by a self-adjoint operator whose eigenvalues are the possible output values, and whose eigenvectors are called the pure states.  As a result, a typical quantum model simply consists of linear operators which form an algebra.

However, when passing from the quantum world to the real world, an actual probabilistic projection to a pure state is necessary.  Such a probabilistic operation destroys the linear structure, so we need a nonlinear (meaning non-distributive) algebraic structure to accommodate such operators. In the 20th century, there were several attempts to attack this problem.  See for instance \cite{Jordan1950,Segal1947,Landau-experiment}, and \cite[Chapter 3]{LiebRuhHensch19} for a beautiful and detailed survey. In particular, Pascual Jordan attempted to use near-ring for quantum mechanics.

Let's consider the scenario of quantum computing.

\begin{defn}[\cite{Moore2000QuantumAA}] \label{def:QFA}
	A \textit{quantum finite automata} (QFA) is a tuple $\mathcal{Q}=(V, q_0, F, \Sigma, (U_{\sigma})_{\sigma\in\Sigma})$ where:
	\begin{enumerate}
		\item $V$ is a finite set of states which generate the Hilbert space $\mathcal{H}_V$;
		\item $F\subset V$ is a set of final or accept states;
		\item $q_0$ the initial state which is a unit vector in $\mathcal{H}_V$;
		\item $\Sigma$ is a finite set called the alphabet;
		\item For each $\sigma\in\Sigma$, $U_{\sigma}$ is a unitary operator on $\mathcal{H}_V$.
	\end{enumerate}
\end{defn}

An input to a QFA consists of a word $w$ in the alphabet $\Sigma$ of the form $w=w_1w_2\dots w_n$ where $w_i\in\Sigma$ for all $i$. $w$ acts on the initial state of the QFA by $\langle q_0|U_w$, where $U_w$ is the matrix $U_w:=U_{w_1}U_{w_2}\dots U_{w_n}$, and $\langle q_0|$ is the row vector presentation of $q_0$.  The probability that word $w$ will end in an accept state is \[Pr(w)=\lVert\langle q_0|U_wP\rVert^2\] 
where $P:\mathcal{H}_V\to\mathcal{H}_F$ is the projection from $\mathcal{H}_V$ to subspace $\mathcal{H}_F$ spanned by $F$. 

Note that the above definition has not taken the probabilistic projection into account.  We make the following reformulation.  
\begin{defn} \label{def:qcomp}
A \textit{quantum computing machine} is a tuple $((\mathcal{H}_V,h),\mathcal{H}_F,e,\rho_G,\sigma^F)$, where
\begin{enumerate}
	\item $(\mathcal{H}_V, h)$ is a Hermitian vector space; \item $\mathcal{H}_F=\mathbb{C}^n$ equipped with the standard metric, which is called the framing space;
	\item $e: \mathcal{H}_F \to \mathcal{H}_V$ is an isometric embedding;
	\item $\rho_G: G \to U(\mathcal{H}_V,h)$ is a unitary representation of a group $G$.
	\item $\sigma^F: \mathcal{H}_F \to \mathcal{H}_F$ is a probabilistic projection.
\end{enumerate}
\end{defn}
Ignoring the last item (5) for the moment, this coincides with Definition \ref{def:QFA} by setting $G=\langle \Sigma \rangle$, the free group generated by a set $\Sigma$, and fixing an initial vector $q_0 \in \mathcal{H}_V$.

Here, we treat $\mathcal{H}_F$ as a vector space of its own and take an isometric embedding $e:\mathcal{H}_F \to \mathcal{H}_V$, rather than directly identifying $\mathcal{H}_F$ as a subspace in $\mathcal{H}_V$.  The state space $\mathcal{H}_V$ is treated as an abstract vector space without a preferred basis, while $\mathcal{H}_F$ is equipped with a fixed basis that has a real physical meaning (like up/down spinning of an electron).  The framing map $e:\mathcal{H}_F \to \mathcal{H}_V$ is interpreted as a bridge between the classical and the quantum world; the image of the fixed basis under $e$ determines a subset of pure state vectors of a certain observable.  The adjoint $e^*: \mathcal{H}_V \to \mathcal{H}_F$ is an orthogonal projection.  In the next section, $e$ is no longer required to be an embedding when we consider \emph{non-unitary generalizations for machine learning}.

For the last item (5), the probabilistic projection $\sigma^F:\mathcal{H}_F\to \mathcal{H}_F$ can be modeled by a probability space.  Namely, consider a $\mathcal{H}_F$-family of random variables 
$$k: \mathcal{H}_F\times \Omega \to \{1,\ldots, |F| \}$$
where $\Omega$ is a probability space (that has a probability measure), with the assumption that $\mathrm{Pr}(k(v) = j) = \langle \frac{v}{\|v\|}, \epsilon_j\rangle$ for every $v\in \mathcal{H}_F$, where $\epsilon_j \in \mathcal{H}_F$ denotes the $j$-th basic vector.  Then $\sigma^F(v) := \epsilon_{k(v)}$.

The major additional ingredients in Definition \ref{def:qcomp}, compared to \ref{def:QFA}, are $e, e^*$ and $\sigma^F$.  Note that they are not yet included in the machine language, which is currently the group $G$.
Since $e$ and $\sigma^F$ are not invertible, we cannot enlarge $G$ to include $e$ nor $\sigma^F$  as a group.

To remedy this, first note that (4) can be replaced with an \emph{algebra} rather than a group, which exhibits linearity and allows not being invertible.  Namely we require instead:
\begin{enumerate}
	\item[(4')] $\rho_A: A \to \End (\mathcal{H}_V)$ is an algebra homomorphism for an algebra $A$ (with unit $1_A$).
\end{enumerate}
For instance, $A$ can be the free algebra generated by a set $\Sigma$.  

With such a modification, we can easily include the framing $e$ and $e^*$ into our language by taking the augmented algebra 
\begin{equation} \label{eq:cA}
\mathcal{A}=A\langle 1_F, e, e^*\rangle / R
\end{equation}
where $R$ is generated by the relations $1_F \cdot 1_F = 1_F, 1_F \cdot e^* = e^*, e \cdot 1_F = e, 1_A \cdot e = e, e\cdot e = 0, e^* \cdot e^* =0, 1_F \cdot e = 0, 1_F \cdot a = 0, e \cdot 1_A =0, e^* \cdot 1_F =0, 1_A \cdot e^*=0$ for any $a \in A$.  The unit of $\mathcal{A}$ is $1_A + 1_F$.

However, we cannot further enlarge $\mathcal{A}$ to include $\sigma^F$ as an algebra.  The reason is that $\sigma^F$ always maps to unit vectors and cannot be linear:
\[\sigma^F(v+w)\neq\sigma^F(v)+\sigma^F(w). \]

To extend $\mathcal{A}$ by $\sigma^F$ which models actual quantum measurement, we need the notion of a \emph{near-ring}.  It is a set $A$ with two binary operations $+$, $\circ$ such that $A$ is a group under `$+$', 
`$\circ$' is associative, and right multiplication is distributive over addition: $(x+y)\circ z = x\circ z + y\circ z$ for all $x, y, z \in A$
(but left multiplication is not required distributive: $z \circ (x+y) \not= z \circ x + z \circ y$).  Near-algebras was introduced by \cite{Brown-thesis}, and the analysis of normed near-algebras was studied in \cite{Irish-thesis}.

Define $\widetilde{\mathcal{A}}$ to be the near-ring \[\widetilde{\mathcal{A}}:=(1_F+e^*\cdot\mathcal{A}\cdot e)\{\sigma^F\}.\] This near-ring can be understood as the language that controls quantum computing machines. Elements of $\widetilde{\mathcal{A}}$ can be recorded as rooted trees.  An example is $a_1 \sigma^F \circ (a_{11} + a_{12})$ where $a_1,a_{11},a_{12} \in (1_F+e^*\cdot\mathcal{A}\cdot e)$.  See also the tree on the left hand side of Figure \ref{fig:example-tree}.

The advantage of putting all the algebraic structures into a single near-ring is that, we can consider all the quantum computing machines (mathematically $\tilde{\mathcal{A}}$-modules) controlled by a single near-ring at the same time.  An element of $\tilde{\mathcal{A}}$ is a quantum algorithm, which can run in all quantum computers controlled by $\tilde{\mathcal{A}}$. 

\section{Near-ring and differential forms}

In the setting of Definition \ref{def:QFA} and Example \ref{ex:differential}, it is natural to relax the representations from unitary groups to matrix algebras $\mathfrak{gl}(n,\mathbb{C})$.  Moreover, the quantum measurement can also be simulated by a non-linear function (called activation function).  Such a modification will produce a computational model of deep learning.

\begin{defn}
	An activation module consists of:
	\begin{enumerate}
		\item A noncommutative algebra $A$ and vector space $V$, $F$;
		\item A family of metrics $h_{(\rho, e)}$ on $V$ over the space of framed $A$-modules \[R=\text{Hom}_{\text{alg}}(A, \text{End}(V))\times\text{Hom}(F, V)\] which is $G$-equivariant where $G=\GL(V)$;
		\item A collection of possibly non-linear functions \[\sigma_j^F:F\to F.\]
	\end{enumerate}
\end{defn}

As in \eqref{eq:cA}, we take the augmented algebra $\mathcal{A}=A\langle 1_F, e, e^*\rangle / R$ which produces linear computations in all framed $A$-modules simultaneously.
With item (3), elements in the near ring $$\tilde{\mathcal{A}} := (1_F + e^*\cdot \mathcal{A}\cdot e)\{\sigma^F_1,\ldots,\sigma^F_N\}$$ 
induce non-linear functions on $F$, and so they are called \textit{non-linear algorithms}.  An example of how $\widetilde{\mathcal{A}}$ induces non-linear functions on $F$ upon fixing a point in $R$ is given in Example \ref{ex:differential}.

$R=\text{Hom}_{\text{alg}}(A, \text{End}(V))\times\text{Hom}(F, V)$ is understood as a family of computing machines: a point $(w,e)$ in $R$ fixes how $\mathcal{A}$ acts on $V$ and the framing map $e\in \Hom(F,V)$, and hence entirely determines how an algorithm runs in the machine corresponding to $(w,e)$.

Let us emphasize that the state space $V$ is basis-free.  The family of metrics is $\GL(V)$-equivariant: $h_{(\rho,e)}(v,w)=h_{g\cdot (\rho,e)}(g\cdot v, g \cdot w)$ for any $g \in \GL(V)$.  Thus, given $a \in \tilde{\mathcal{A}}$, the non-linear functions that $a$ induces for the two machines $r \in R$ and $g\cdot r \in R$ equal to each other.  In other words, an algorithm $a \in \tilde{\mathcal{A}}$ drives all machines parametrized by the \emph{moduli stack} $[R/\GL(V)]$ to produce functions on $F$:
\begin{equation} \label{eq:Map}
\tilde{\mathcal{A}} \times [R/\GL(V)] \to \textrm{Map}(F,F).
\end{equation}
As mentioned above, the advantage is that the single near-ring $\tilde{A}$ controls all machines in $[R/\GL(V)]$ and for all $V$ simultaneously (independent of $\dim V$).  

In \cite{JL22}, we formulated noncommutative differential forms on a near-ring $\tilde{\mathcal{A}}$, which induce $\textrm{Map}(F,F)$-valued differential forms on the moduli $[R/\GL(V)]$.  It is extended from the Karoubi-de Rham complex \cite{Connes}\cite{CQ}\cite{Ginzburg-quiver}\cite{Tacchella} for algebras to near-rings.
\eqref{eq:Map} above is the special case for $0$-forms, which are simply elements in $\tilde{\mathcal{A}}$.  The cases of $0$-forms and $1$-forms are particularly important for gradient descent: recall that gradient of a function is the metric dual of the differential of that function.

\begin{theorem}[\cite{JL22}] \label{thm:Atildeform}
	There exists a degree-preserving map $$DR^\bullet(\tilde{\mathcal{A}}) \to  (\Omega^\bullet(R,\mathrm{Map}(F,F)))^{\GL(V)}$$
	which commutes with $d$ on the two sides.

\end{theorem}

	A differential form on $\tilde{\mathcal{A}}$ can be recorded as a \textit{form-valued tree}, see the right hand side of Figure \ref{fig:example-tree}.  They are rooted trees whose edges are labeled by $\phi\in DR^{\bullet}(\text{Mat}_F(\hat{\mathcal{A}}))$; leaves are labeled by $\alpha\in\widetilde{\mathcal{A}}$; the root is labeled by 1 (if not a leaf); nodes which are neither leaves nor the root are labeled by the symbols $D^{(p)}_{\sigma_{\ell}}|_{\alpha}$ that correspond to the $p$-th order symmetric differentials of $\sigma_{\ell}$.
	

In application to machine learning, an algorithm $\tilde{\gamma} \in \widetilde{\mathcal{A}}$ induces a 0-form of $\widetilde{\mathcal{A}}$, for instance
\begin{equation} \label{eq:integrate}
\int _{K}\left| \tilde{\gamma}(x)-f\left(x\right)\right|^{2} dx
\end{equation}
for a given dataset encoded as a function $f: K \to \R$.  This 0-form and its differential induces the cost function and its differential on $[R/G]$ respectively, which are the central objects in machine learning.

The differential forms are $G$-equivariant by construction.  There have been a lot of recent works in learning for input data set that has Lie group symmetry \cite{Barbaresco21, CW, CGW, CWKW-spherical, CWKW, CAWHCW, dehaan2020natural}.  On the other hand, our work has focused on the internal symmetry of the computing machine.


In general, the existence of fine moduli is a big problem in mathematics: the moduli stack $[R/G]$ may be singular and pose difficulties in applying gradient descent.  Fortunately, if $A$ is a quiver algebra, its moduli space of framed quiver representations $[R/G]$ is a smooth manifold $\mathcal{M}$ (with respect to a chosen stability condition) \cite{King}.  This leads us to deep learning explained in the next section.

\section{Deep learning over the moduli space of quiver representations}

\begin{figure}
	\centering
	\includegraphics[scale=0.3]{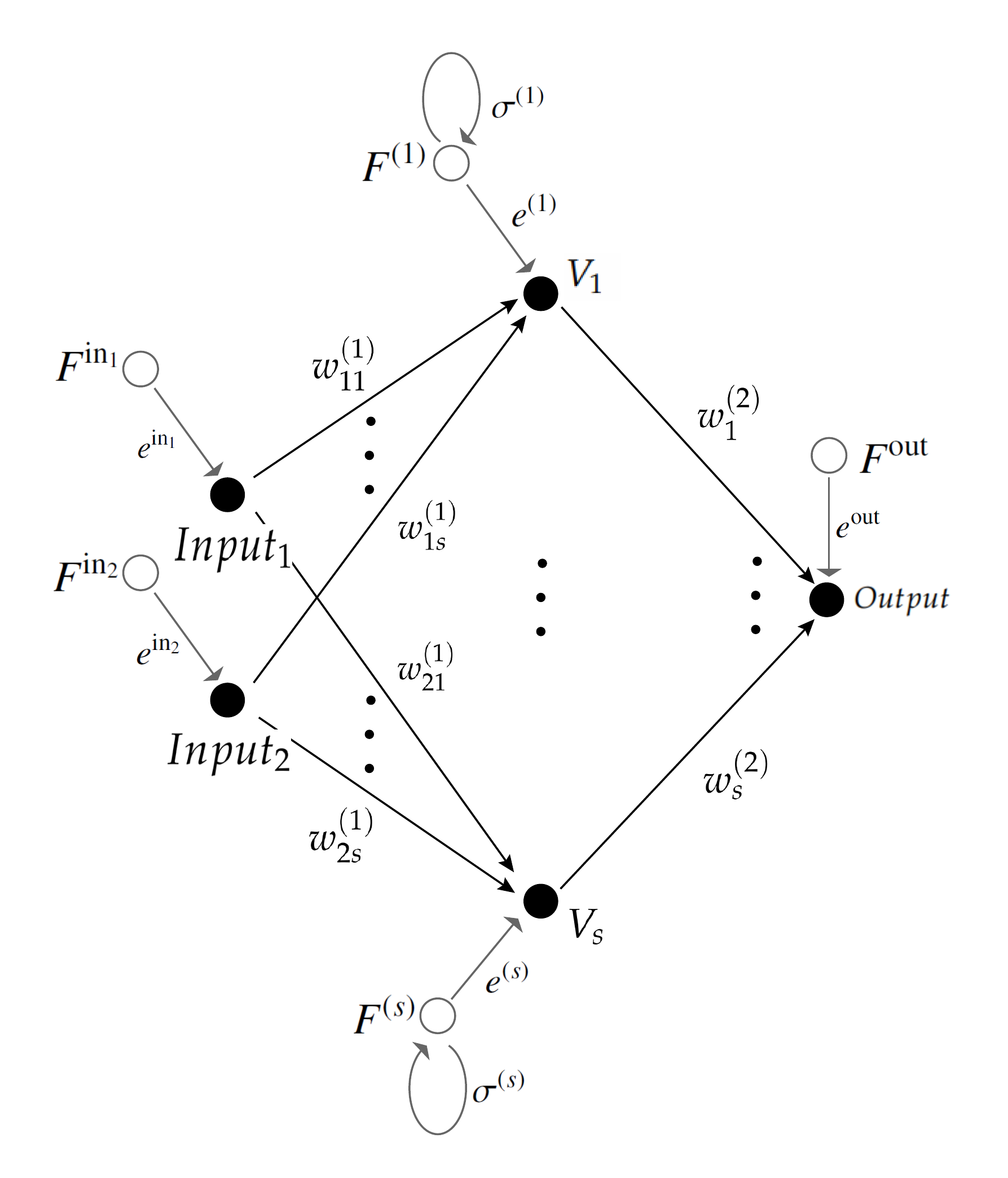}
	\caption{A simple neural network with one hidden layer with $s$-many neurons. 
	Neural networks in applications are typically much more complicated, but in nature are still quiver representations equipped with activation functions.}
	\label{fig:network}
\end{figure}

An artificial neural network (see Figure \ref{fig:network} for a simple example) consists of:
\begin{enumerate}
	\item a \emph{graph} $Q=(Q_0, Q_1)$, where $Q_0$ is a (finite) set of vertices (neurons) and $Q_1$ is a (finite) set of arrows starting and ending in vertices in $Q_0$ (transmission between neurons);
	\item a \emph{quiver representation} of $Q$, which associates a vector space $V_i$ to each vertex $i$ and a linear map $w_a$ (called weights) to each arrow $a$.  We denote by $t(a)$ and $h(a)$ the tail and head of an arrow $a$ respectively.
	\item a non-linear function $V_i \to V_i$ for each vertex $i$ (called an \emph{activation function} for the neuron).
\end{enumerate}
Activation functions are an important ingredient for neural network; on the other hand it rarely appears in quiver theory.  Its presence allows the neural network to produce non-linear functions. 

\begin{figure}[h]
\centering
\includegraphics[scale=0.12]{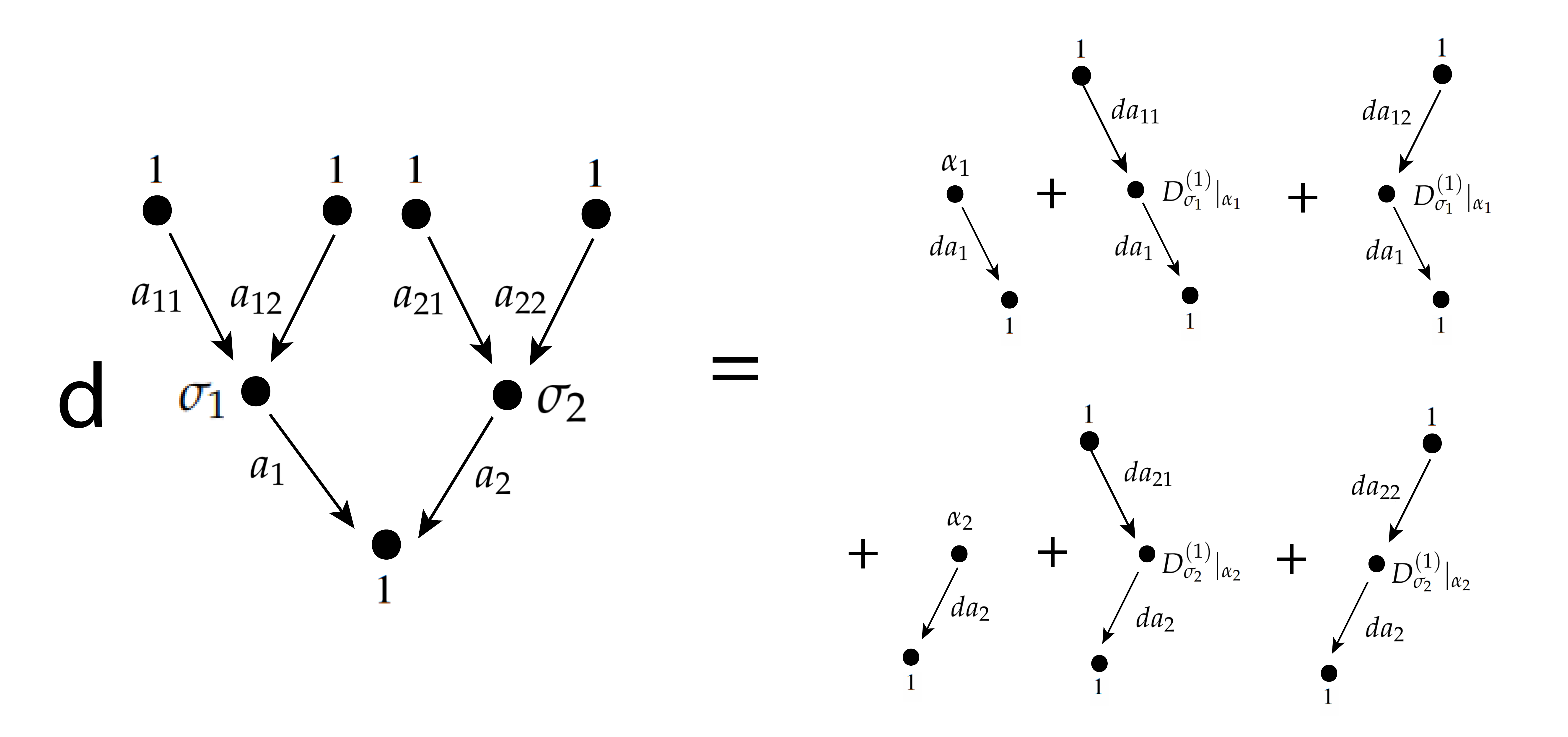}
\caption{An example of the backpropagation algorithm for the network in Figure \ref{fig:network} when $s=2$.}
\label{fig:example-tree}
\end{figure}

\begin{remark}
	In the recent past there has been rising interest in the relations between machine learning and quiver representations \cite{Armenta-Jodoin, JL21, GW, JL22}.  Here, we simply put quiver representation as a part of the formulation of an artificial neural network.  

	In many applications, the dimension vector $\vec{d} \in \Z_{\geq 0}^{Q_0}$ is set to be $(1,\ldots,1)$, that is, all vector spaces $V_i$ associated to the vertices are one-dimensional.  For us, it is an unnecessary constraint and we allow $\vec{d}$ to be any fixed integer vector.
\end{remark}

Any non-trivial non-linear function $V_i\to V_i$ cannot be $\GL(V_i)$-equivariant.  However, in quiver theory, $V_i$ is understood as a basis-free vector space and requires $\GL(V_i)$-equivariance.  We resolve this conflict between neural network and quiver theory in \cite{JL21} by using framed quiver representations.  The key idea is to put the non-linear function on the framing rather than on the basis-free vector spaces $V_i$.

Combining with the setting of the last section (Definition \ref{def:qcomp}), we take:
\begin{enumerate}
	\item $A = \mathbb{C} Q$, the quiver algebra.  Elements are formal linear combinations of paths in $Q$ (including the trivial paths at vertices), and product is given by concatenation of paths.
	\item $V = \bigoplus_i V_i$, the direct sum of all vector spaces over vertices.
	\item Each vertex is associated with a framing vector space $F_i$.  Then $F = \bigoplus_i F_i$.
	\item Each point $(w,e) \in R=\text{Hom}_{\text{alg}}(\mathbb{C} Q, \text{End}(V))\times\bigoplus_i \text{Hom}(F_i, V_i)$ is a framed quiver representation.  Namely, $w \in \text{Hom}_{\text{alg}}(\mathbb{C} Q, \text{End}(V))$ associates a matrix $w_a$ to each arrow $a$ of $Q$; $e^{(i)} \in \text{Hom}(F_i, V_i)$ are the framing linear maps.   
	\item The group $G$ is taken to be $\prod_i \GL(V_i)$.  An element $g=(g_i)_{i\in Q_0}$ acts on $R$ by
	\[g\cdot \left((w_a)_{a\in Q_1}, (e^{(i)})_{i \in Q_0}\right)=\left((g_{h(a)}\cdot w_a \cdot g_{t(a)}^{-1})_{a\in Q_1}, (g_i\cdot e^{(i)})_{i\in Q_0} \right). \] 
	\item We have (possibly non-linear) maps $\sigma_i: F_i \to F_i$ for each vertex.  To match the notation of Definition \ref{def:qcomp}, $\sigma_i$ can be taken as maps $F \to F$ by extension by zero.
\end{enumerate}

By the celebrated result of \cite{King}, we have a fine moduli space of framed quiver representations $\mathcal{M} = \mathcal{M}_{n,d} = R\sslash G$, where $n,d$ are the dimension vectors for the framing $\{F_i\}_{i\in Q_0}$ and representation $\{V_i\}_{i\in Q_0}$ respectively.  In particular, we have the universal vector bundles $\cV_i$ over $\mathcal{M}$, whose fiber over each framed representation $[w,e]\in \mathcal{M}$ is the representing vector space $V_i$ over the vertex $i$.  

$\cV_i$ plays an important role in our computational model, namely, a vector $v \in \mathcal{V}_i$ over a point $[w,e]\in \mathcal{M}$ is the state of the $i$-th neuron in the machine parametrized by $[w,e]$.

\begin{remark}
	The topology of $\mathcal{M}$ is well understood by \cite{Reineke} as iterated Grassmannian bundles.  Framed quiver representations and their doubled counterparts play an important role in geometric representation theory \cite{Nakajima-Duke, Nakajima-JAMS}.
\end{remark}

To fulfill Definition \ref{def:qcomp} (see Item (2)), we need to equip each $\mathcal{V}_i$ with a bundle metric $h_i$, so that the adjoint $e^*$ makes sense. In \cite{JL21}, we have found a \emph{bundle metric that is merely written in terms of the algebra $\mathcal{A}$}.  It means the formula works for (infinitely many) quiver moduli for all dimension vectors of representations simultaneously.

\begin{theorem}[\cite{JL21}] \label{theorem:H}
	For a fixed vertex $(i)$, let $\rho_i$ be the row vector whose entries are all the elements of the form $w_{\gamma}e^{(t(\gamma))}:R_{n,d}\to \Hom(\mathbb{C}^{n_{t(\gamma)}}, \mathbb{C}^{d_i})$ such that $h(\gamma)=i$. Consider 
	\begin{equation}\label{equation:metric}
	H_i := \rho_i\rho_i^*=\sum\limits_{h(\gamma)=i}\left(w_{\gamma}e^{(t(\gamma))}\right)\left(w_{\gamma}e^{(t(\gamma))}\right)^*
	\end{equation} 
	as a map $\rho_i\rho_i^*:R_{n,d}\to \text{End}(\mathbb{C}^{d_i})$.	
	Then $(\rho_i\rho_i^*)^{-1}$ is $\GL(d)$-equivariant and descends to a Hermitian metric on $\mathcal{V}_i$ over $\mathcal{M}$.
\end{theorem}


\begin{example}\label{ex:differential}
	
Consider the network in Figure \ref{fig:network}.  The quiver has the arrows $a_{1,k}^{(1)}, a_{2,k}^{(1)}$ for $k=1,\ldots,s$ (between the input and hidden layers) and $a_k^{(2)}$ (between the hidden and output layers).  In application, we consider the algorithm
\[\tilde{\gamma}:=\sum_{k=1}^s \hat{a}^{(2)}_{k}\sigma_k \circ \sum_{j=1}^2 \hat{a}^{(1)}_{jk} \in \tilde{\cA} \] 
where $\hat{a}^{(1)}_{jk} := \left(e^{(j)}\right)^{*}a_{jk}^{(1)}e^{\textrm{in}_j}$ and $\hat{a}^{(2)}_{k} := \left(e^{\textrm{out}}\right)^{*} a_k^{(2)}e^{(k)}$.  Note that the adjoints $\left(e^{(j)}\right)^{*}$ and $\left(e^{\textrm{out}}\right)^{*}$ are with respect to the metric $H_i$ and $H_{\textrm{out}}$ respectively.

 $\tilde{\gamma}$ is recorded by the activation tree on the left hand side of Figure \ref{fig:example-tree}, for the case $s=2$. $\tilde{\gamma}$ drives any activation module (with this given quiver algebra) to produce a function $F_{\mathrm{in}_1} \times F_{\mathrm{in}_2} \to F_{\mathrm{out}}$.  For instance, setting the representing dimension to be $1$ and taking $\sigma_i$ to be the ReLu function $\max\{x, 0\}$ on $\R$ is a popular choice. 
 Data passes from the leaves to the root, which is called forward propagation.


Figure \ref{fig:example-tree} shows the differential of $\tilde{\gamma} \in \tilde{\cA}$.  This 1-form is given by
\[d(a_1 \sigma_1 \circ (a_{11}+a_{12})+a_2\sigma_2 \circ (a_{21}+a_{22}))\]\[=da_1(\alpha_1+D_{\sigma_1}^{(1)}|_{\alpha_1}(da_{11}+da_{12}))+da_2(\alpha_2+D_{\sigma_2}^{(1)}|_{\alpha_2}(da_{21}+da_{22}))\]
where $\alpha_j = \sigma_j \circ (a_{j1}+a_{j2})$.  The terms are obtained by starting at the output node and moving backwards through the activation tree, which is well known as the backpropagation algorithm.  Note that this works in the algebraic level and is not specific to any representation.

$d\tilde{\gamma}$ induces a $\textrm{Map}(F_{\mathrm{in}_1} \times F_{\mathrm{in}_2}, F_{\mathrm{out}})$-valued 1-form on $\mathcal{M}_{(n, d)}$.  We can also easily produce $\R$-valued 1-forms, for instance by \eqref{eq:integrate}.  
\end{example}



For stochastic gradient descent over the moduli $\mathcal{M}_{n,d}$ in order to find the optimal machine, we still need one more ingredient: a metric on $\mathcal{M}_{n,d}$, to turn a one-form to a vector field.  Very nicely, the Ricci curvature of the metric \eqref{equation:metric} given above gives a well-defined metric on $\mathcal{M}_{n,d}$.  So far, all the ingredients involved (namely, the algorithm $\tilde{\gamma}$, its differential, the bundle metric $H_i$ and the metric on moduli) are purely written in algebraic symbols and work for moduli spaces in all dimensions $(n,d)$ simultaneously.

\begin{theorem}[\cite{JL21}] \label{theorem:metric}
	Suppose $Q$ has no oriented cycles. Then
	\begin{equation} \label{eq:H_T}
	H_T:=\sum_i\partial\bar{\partial}\log \textrm{det } H_i =\sum_i\left(tr(\partial\rho_i)^*H_i\partial\rho_i-tr\left(H_i\rho_i(\partial\rho_i)^*H_i(\partial\rho_i)\rho_i^*\right)\right)
	\end{equation}
	defines a K\"ahler metric on $\mathcal{M}_{n,d}$ for any $(n,d)$.
\end{theorem}

\begin{example}
	Let's consider the network of Figure \ref{fig:network} again, with $s=2$ for simplicity.  Let $n=d=(1, 1, 1, 1)$. Over the chart where $e^{(i)} \not=0$ for all $i=\textrm{in}_1, \textrm{in}_2, 1,2, \textrm{out}$, the $\GL(d)$-equivariance allows us to assume that $e^{(i)}(e^{(i)})^*=1$. Then $H_{\textrm{in}_j}$ are trivial for $j=1,2$, and so $\partial\bar{\partial}\log H_{\textrm{in}_j}=0$.  Let $x_1=(w_{11}^{(1)}, w_{21}^{(1)})$ and $x_2=(w_{12}^{(1)}, w_{22}^{(1)})$.  We have \[\partial\bar{\partial}\log H_j=\partial\bar{\partial}\log\left(1+|x_j|^2\right)^{-1}=\frac{(1+|x_j|^2)dx_j\wedge d\bar{x}_j^t+\bar{x_j}dx_j^td\bar{x_j}x_j^t}{(1+|x_j|^2)^2};\]
\[\partial\bar{\partial}\log H_{out}=\partial\bar{\partial}\log\left(1+|w_1^{(2)}|^2|x_1|^2+|w_2^{(2)}|^2|x_2|^2\right)^{-1}=\frac{dx_j\wedge d\bar{x_j}+d w_j^{(2)}\wedge d\bar{w_j}^{(2)}}{(1+|w_1^{(2)}|^2|x_1|^2+|w_2^{(2)}|^2|x_2|^2)}\]\[+\frac{(|w_j^{(2)}|^2\bar{x_j}dx_j^td\bar{x_j}x_j^t)+(|x_j|^2dw_j\wedge d\bar{w_j})+(\bar{w_j}x_jdw_j\wedge d\bar{x_j})+(w_j\bar{x_j}dx_j\wedge d\bar{w_j})}{(1+|w_1^{(2)}|^2|x_1|^2+|w_2^{(2)}|^2|x_2|^2)^2}.\]
	

\end{example}

\section{Uniformization of metrics over the moduli}\label{sec:uniform}

The original formulation of deep learning is over the flat vector space of representations \newline $\text{Hom}_{\text{alg}}(\mathbb{C} Q, \text{End}(V))$, rather than the moduli space $\mathcal{M} = R\sslash G$ of framed representations which has a semi-positive metric $H_T$.  In \cite{JL22}, we found the following way of connecting our new approach with the original approach by varying the bundle metric $H_i$ in \eqref{equation:metric}.

We shall assume $n_i\geq d_i$ $\forall i$.  Let's write the framing map (which is a rectangular matrix) as $e^{(i)}=(\epsilon_i$ $b_i)$ where $\epsilon_i$ is the largest square matrix and $b_i$ is the remaining part.  (In applications $b_i$ usually consists of `bias vectors'.)  This allows us to rewrite Equation \ref{equation:metric} in the following way:
\[
H_i(\alpha)=\left(\epsilon_i\epsilon_i^*+\alpha_{e^{(i)}} b_i b_i^* + \sum_{\gamma:h(\gamma)=i, \gamma\neq e^{(i)}}\alpha_{\gamma}w_{\gamma}e^{t(\gamma)}\left(w_{\gamma}e^{t(\gamma)}\right)^*\right)^{-1}.
\]
with $\alpha = (\alpha_\gamma)_{\gamma: h(\gamma)=i} = (1,\ldots,1)$.

If instead, we set $\alpha_{\gamma}$ to different values, then $H_i(\alpha)$ will still be $G$-equivariant, but it may no longer positive definite on the whole space $\mathcal{M}$.  Motivated by the brilliant construction of dual Hermitian symmetric spaces, we define
$$ \mathcal{M}(\alpha) := \{[w,e] \in \mathcal{M}: H_i(\alpha) \textrm{ is positive definite at } [w,e]\}.$$
Elements $[w,e] \in  \mathcal{M}(\alpha)$ are called space-like representations with respect to $H_i(\alpha)$.
If we set $\alpha = \vec{0}$, it becomes
\[H_i^0:=\left(\epsilon_i\epsilon_i^*\right)^{-1}\]
and $\mathcal{M}^0 := \mathcal{M}(\alpha=\vec{0})$ is exactly the flat vector space $\text{Hom}_{\text{alg}}(\mathbb{C} Q, \text{End}(V))$.   This recovers the original Euclidean learning.

On the other hand, setting $\alpha = (-1,\ldots,-1)$, we obtain a semi-negative moduli space $(\mathcal{M}^-,H_T)$ \cite{JL22}, which is a generalization of the hyperbolic spaces or non-compact dual of Grassmannians. This is a very useful setting as there have been several fascinating works done on machine learning performed over hyperbolic spaces, for example  \cite{Nickel2017PoincarEF, Ganea2018HyperbolicEC, Sala2018RepresentationTF, Ganea2018HyperbolicNN}.

Thus we have a family of metrics parametrized by $\alpha$.  These parameters $\alpha_\gamma$ can be optimized during the learning algorithm.  It is interesting to compare with the celebrated attention mechanism.  Namely, $\alpha_\gamma$ can be interpreted as learning parameters that encode the importance of the paths $\gamma$.

\section*{Acknowledgment}

We express our deep gratitude to Bernd Henschenmacher for the very useful discussions and references on the historical attempts of using near-rings and near-algebras in quantum physics.  They gave us a lot of motivations and encouragements to deepen the study in this direction.  The project is partially supported by Simons Collaboration Grant.

\bibliographystyle{amsalpha}
\bibliography{geometry}	

\end{document}